# An Introduction to Knowledge Management

Sabu M. Thampi {smtlbs@gmail.com}
L.B.S Institute of Technology for Women, Trivandrum, Kerala, India

## Abstract

*Knowledge has been lately recognized as one of the most important assets of organizations. Managing knowledge has grown to be imperative for a company's success. This paper presents an overview of Knowledge Management and various aspects of secure knowledge management. A case study of knowledge management activities at Tata Steel is also discussed.*

## 1. Introduction

Following technological advancement, machinery came into picture of production and it began to perk up the automation, and industry had no longer to rely on labours largely. However, as a result of huge investment in machinery, capital turned out to be more powerful. Managing flow of capital was a most important problem for the industrialists and swiftly capital became the bottleneck to competence. Though conventional three issues of production – land, labour and capital – have become easier to handle in 21st century, a fourth issue is more and more becoming a barrier for companies to grow. This is "knowledge", which is at the heart of much of today's global economy and managing knowledge has grown to be imperative for company's success. Knowledge has been recently recognized as one of the most significant resources of organizations. The capturing of an organization's knowledge facilitates an organization remain competitive and can help an organization continue to exist in the business world of today. 'Knowledge' can be defined as a fluid mix of experience, values, contextual information and expert insight that offers a framework for assessing and incorporating new experiences and information. In ancient Greece, the philosopher, Plato, in his dialogues, captured and elaborated the thoughts of his mentor Socrates, and as a result, succeeding generations have been able to discover and share that thoughts, and in turn reinterpret those thoughts and to be stimulated to accomplish fresh insights and creativity.

The following bank account example illustrates how data, information, and knowledge relate to principal, interest rate, and interest. The numbers 100 or 5%; completely out of context, are just pieces of *data*. Interest, principal, and interest rate, out of context, are not much more than data as each has multiple meanings which are context dependent. If a bank savings account as the basis for context is established, the interest, principal, and interest rate become meaningful (*information*) in that situation with specific interpretations: Principal is the amount of money-Rs. 100, in the savings account, Interest rate- 5%, is the factor used by the bank to compute interest on the principal. If Rs.100 is deposited in the savings account, and the bank pays 5% interest yearly, then at the end of one year the bank will compute the interest of Rs. 5 and add it to the principal. Subsequently, the account will have Rs. 105 in the bank. This pattern corresponds to *knowledge*, which permits the account holder to know how the pattern will evolve in time and the results it will generate. In understanding the pattern, what the customer is familiar with is knowledge - if the customer deposits more money in the account, he/she will receive more interest, while if the customer withdraws cash from the account, he/she will earn less interest. Thus, knowledge is information in action and it is what people in an organization know about their customers, products, processes, mistakes, and successes. Unlike the conventional material assets, which diminish as they are used, knowledge asset augments with use, and shared knowledge stays with the contributor while it enriches the receiver.

*Knowledge management (KM)* is the name given to the set of systematic and regimented actions that an organization can take to attain the maximum value from the knowledge available to it. Effective knowledge management normally requires a proper amalgamation of organizational, social, and managerial initiatives along with exploitation of apposite technology. The idea of KM is to gather, classify, store and spread all knowledge that is needed to make the organization both grow and flourish. The focus of KM is to leverage and reuse knowledge resources that previously subsist in the organization so that people will seek out best practices rather than reinvent the wheel. Knowledge management deals with two types of knowledge, *tacit* and *explicit*. In effect, these two types of knowledge are like two sides of the same coin, and are equally pertinent for the overall knowledge of an organization. Explicit knowledge (sometimes referred to as *formal knowledge*) is formal knowledge that can be packaged as information. For example, it is very simple to instruct someone how to calculate the area of a rectangle, a familiar example of explicit knowledge, by indicating that the area is a result of calculating the length of the rectangle by the width of the rectangle. Explicit knowledge can be found in an organization in the form of reports, articles, manuals, patents, pictures, images, video, sound, software etc., which have been created with the goal of communicating with another person. Explicit knowledge defines the identity, the competencies and the intellectual assets of an organization independently of its employees; thus, it is organizational knowledge *par excellence*, but it can grow and sustain itself only through an affluent background of tacit knowledge.

Tacit knowledge (also, *informal knowledge*) is personal knowledge embedded in individual experience and is shared and exchanged through direct, eye-to-eye contact. Tacit knowledge is knowledge the "knower" is not aware of. Individuals may not know what tacit knowledge they have and also might not be able or willing to externalize it. For example, it is very difficult to explain how to drive a car. It takes experience to really know how to adjust the pressure difference one must put on the gas petal when driving from a flat road to a hill or how much to turn the wheel when changing lanes. Tacit knowledge can only be elicited, and thus articulated, with great effort and the use of special observation or interview techniques. Tacit knowledge is much more difficult to manage. With experience and continued learning, the tacit knowledge matures and evolves into new knowledge, which remains tacit within the individual or group until they document it in some fashion, making it explicit. Both forms of knowledge are important for organizational effectiveness. In simple language KM is an effort to capture not only explicit factual information but also the tacit information and knowledge that exists in an organization, usually based on the experience and learning of individual employees, in order to advance the organization's mission. The eventual goal is to share knowledge among members of the organization.

Recently, the use of information technologies within an organization has been identified, by many companies, as an important tool for managing or sharing organizational knowledge in order to improve business performance. The value of knowledge to an organization is supported with the growing support and implementation of knowledge management systems. Knowledge management systems are designed to help organizations capture, codify, store, and dissemination organizational knowledge. Knowledge management systems come in many forms such as expert systems and knowledge management portals. All knowledge management systems, however, must have safeguards in place for protecting an organization's knowledge.

## 2. Knowledge Management at TATA Steel

Tata Steel [3] is the world's 6th largest steel company with an existing annual crude steel production capacity of 30 million tones per annum. Established in 1907, it is the first integrated steel plant in Asia and is now the world's second most geographically diversified steel producer and a Fortune 500 Company. Tata Steel has a balanced global presence in over 50 developed European and fast growing Asian markets, with manufacturing units in 26 countries.

Tata Steel decided to embark on formal KM initiative in the year 1999. The beginning was made in July'99 to place a Knowledge Management programme for the company to systematically and formally share and transfer learning concepts, and best practices. The various phases of KM spiral at Tata Steel are shown in Figure 1. The Knowledge Management has been identified as one of the main enablers to make Tata Steel self reliant in technology. Tata Steel aims at capturing knowledge from various working groups and outside agencies. The major stakeholders covered under KM being senior management, officers, employees, customers, suppliers and experts in and outside company. On the corporate intranet a KM Portal has been developed to communicate all KM related matters across the company. It provides an online knowledge repository to the users who can submit, search and use knowledge pieces available on it. The portal also provides a virtual forum where employees can invite and involve other fellow employees or lead experts to discuss and solve the problems faced by them. Relevant Indian and international standards, quality system manuals, standard practices and procedures also feature for ready reference of users.

Tata Steel follows three strategies for managing organizational knowledge. Knowledge can be contributed either by an individual (codification) or a team or a group of people (personalization). The first two strategies enable capture and systematic storage of knowledge, whereas the third strategy (knowledge diffusion) derives the benefit of replicating best practices identified in the repository and thereby eliminating the 're-invention of wheel'. The above strategies ensure knowledge sharing across the entire value chain from customer to the supplier. Tata Steel has been recognized as the overall (1st place) 2006 Indian Most Admired Knowledge Enterprises (MAKE) Winner compared to its 6th position for the year 2005.

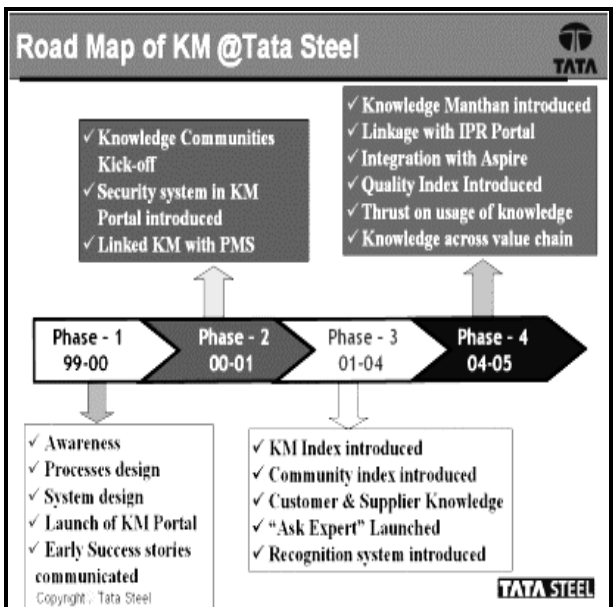

**Figure 1: Various Phase of KM at Tata Steel [3]**

## 3. Secure Knowledge Management

With the advent of intranets and web-access, it is even more crucial to protect corporate knowledge as numerous individuals now have access to the assets of a corporation. Security is becoming a major issue revolving around Knowledge Management Systems (KMS). Security methods [1, 2, 5] may include authentication or passwords, cryptography programs, intrusion detection systems or access control systems. Issues include protecting from malicious insiders, securing against subversion attacks and establishing correct policies and refinement and enforcement. Furthermore KMS content is much more sensitive than raw data stored in databases and issues of privacy also become important.

Security strategies for knowledge management include the policies and procedures that an organization sets in place for secure data and information sharing as well as protecting the intellectual property. Security has to be incorporated into the business processes for workflow, contracting, and purchasing. For example, only users with certain credentials can carry out various knowledge-management processes. Metrics for secure knowledge management should focus on the impact of security on knowledge-management metrics. Some examples of knowledge-management metrics include the number of documents published, number of conferences attended, or the number of patents obtained. When security is incorporated, then the number of documents published may decrease as some of the documents may be classified. Organizations should carry out experiments determining the impact of security on the metrics gathered. Security techniques include access control, trust management, as well as privacy control. These techniques are enforced at all stages of knowledge-management processes. Knowledge management includes many technologies such as data mining, multimedia, collaboration, and the web. The component technologies have to be secure if we are to ensure secure knowledge management.

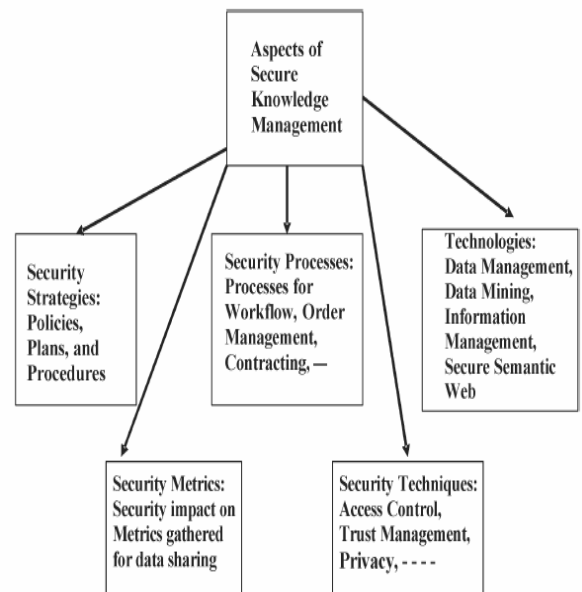

**Figure 2: Aspects of secure knowledge management [1]**

## 4. Conclusions

A short introduction to knowledge management is given, and also an example for a knowledge-management system is briefly explained. Moreover the important features of secure knowledge management systems are discussed. Secure knowledge management will continue to be critical as organizations work together, share data, as well as collaborate on projects. Remember that "Processing data can be performed by machine, but only the human mind can process knowledge or even information".